\documentclass[10pt,journal,jrnl,twocolumn]{IEEEtran}

\usepackage{booktabs} 
\usepackage{multirow}
\usepackage{flushend}
\usepackage{subfigure}
\usepackage{hyperref}
\usepackage{dsfont,amsfonts,amssymb,amsmath,color} 
\usepackage{latexsym}
\usepackage{graphicx}
\usepackage{cleveref}
\usepackage{enumitem}
\usepackage[ruled]{algorithm2e}
\usepackage{indentfirst}

\begin{document}

\title{FedMood: Federated Learning on Mobile Health Data for Mood Detection}

\author{
	\IEEEauthorblockN{Xiaohang Xu\IEEEauthorrefmark{1},
		Hao Peng\IEEEauthorrefmark{1},\IEEEauthorrefmark{2},
		Lichao Sun\IEEEauthorrefmark{3},
		Md Zakirul Alam Bhuiyan\IEEEauthorrefmark{4},
		Lianzhong Liu\IEEEauthorrefmark{1},
		Lifang He\IEEEauthorrefmark{5}
	}
	
	\IEEEauthorblockA{\IEEEauthorrefmark{1}School of Cyber Science and Technology, Beihang University, Beijing 100083, China.} 
	
	\IEEEauthorblockA{\IEEEauthorrefmark{2}Beijing Advanced Innovation Center for Big Data and Brain Computing, Beijing 100083, China.} 
	
	\IEEEauthorblockA{\IEEEauthorrefmark{3}Department of Computer Science, University of Illinois at Chicago, Chicago, USA.} 
	
	\IEEEauthorblockA{\IEEEauthorrefmark{4}Department of Computer and Information Sciences Fordham University JMH 334,E Fordham Road, Bronx, NY 10458 USA.}

	\IEEEauthorblockA{\IEEEauthorrefmark{5}Department of Computer Science and Engineering, Lehigh University, Bethlehem, PA 18015 USA.} 
	
}

\maketitle

\maketitle

\begin{abstract}  
Depression is one of the most common mental illness problems, and the symptoms shown by patients are not consistent, making it difficult to diagnose in the process of clinical practice and pathological research.
Although researchers hope that artificial intelligence can contribute to the diagnosis and treatment of depression, the traditional centralized machine learning needs to aggregate patient data, and the data privacy of patients with mental illness needs to be strictly confidential, which hinders machine learning algorithms clinical application.
To solve the problem of privacy of the medical history of patients with depression, we implement federated learning to analyze and diagnose depression. 
First, we propose a general multi-view federated learning framework using multi-source data,which can extend any traditional machine learning model to support federated learning across different institutions or parties.
Secondly, we adopt late fusion methods to solve the problem of inconsistent time series of multi-view data.
Finally, we compare the federated framework with other cooperative learning frameworks in performance and discuss the related results.
\end{abstract}

\begin{IEEEkeywords}
Federated learning; Depression; Data privacy; Mobile device.
\end{IEEEkeywords}

\section{Introduction}\label{sec:intro}

\IEEEPARstart{D}{epression} is a very common disease in real life. 
More than 300 million people worldwide suffer from depression. 

At present, the diagnosis of depression depends almost entirely on the subjective judgment of the doctor through communication with the patient and the relevant questionnaires filled out. 
Hamilton Depression Rating Scale (HDRS) \cite{hamilton1986hamilton} and Young Mania Rating Scale (YMRS) \cite{young1978rating} are commonly used evaluation criteria for doctors when diagnosing depression. 
In order to better help doctors diagnose depression, researchers analyze patient data by introducing machine learning technology \cite{grunerbl2014smartphone}.
But when using machine learning technology, there is a contradiction between the performance of the model and the protection of data privacy \cite{darcy2016machine}.

First, the quality of the model trained by machine learning is closely related to the amount of data.
Deep Neural Network (DNN) has achieved good results in a variety of medical applications, but it highly depends on the amount and diversity of training data \cite{sun2017revisiting}. 
Although Wang et al. \cite{wang2019privacy} proposed a scheme to avoid data privacy leakage under centralized learning, hospitals need to protect the privacy of patients' diagnosis data, so different medical institutions cannot gather and share data \cite{hao2019efficient}, which greatly affects the accuracy of the model \cite{li2019privacy}. 
For example, in the work of electrocardiogram, because a single medical institution cannot collect enough high-quality data, the predictive ability of the model cannot achieve the role of clinical assistance.

Second, although there are many machine learning algorithms that involve privacy protection, it is difficult to achieve good training results.
There are many machine learning application scenarios for privacy protection, such as recommendation systems and face recognition. 
However, the privacy protection machine learning method needs to increase the noise according to the sensitivity of the algorithm's intermediate product, so under the limited privacy budget, the prediction performance of the privacy algorithm is often poor \cite{yao2019privacy}. 
Third, due to the huge gap between various medical institutions, the patient data they have varies greatly. 
In order to deal with various situations, algorithms and software are required to have a high generalization ability, and it is difficult for the model to obtain sufficient accuracy and specificity without data exchange.

To address the above limitations, in 2016, Google \cite{mcmahan2016federated} proposed a method called federated learning to break the problem of medical data silos due to patient data privacy.
Each medical institution does not need to centralize patient data to train a machine learning model. Instead, it aggregates the trained model in one place and uses federated averaging technology to continuously optimize the model so that the data can be available to all health care facilities.
However, most of the existing research using the federated learning framework in the medical field is based on the existing data of hospitals. 
It mainly includes the diagnosis of the characteristics of patients with specific diseases, reducing the cost of diagnosis and treatment, medical image processing and other issues
\cite{li2019privacy}.
As mobile devices become more and more popular, smart phones, bracelets and other devices are also recording users' information all the time.
According to existing researches \cite{agu2013smartphone}, 
as one of the most important tools for information transmission in patients’ lives, mobile phones can also be an important data source for disease prediction.
We believe that keyboard keystrokes, such as the interval between two keystrokes, can be used as a form of biometric identification to predict depression by analyzing the keystroke habits of patients with depression. 
The typing speed of depression patients is usually different from that of normal people, which may be caused by emotional instability during the onset of the disease \cite{hussain2019passive}.

Our work uses a virtual keyboard customized for mobile phones to collect metadata (including key letters, special characters, and phone accelerometer values). 
Using \textgreater1.3 million keypresses from 20 users and each users who additionally completed at least 1 Patient Health Questionnaire. 
We regard the user's keypresses at least five seconds apart as the beginning, and no operation after five seconds of the last keypress as a session carry out. 
The duration of the session is usually within 1 minute. 
We use federated learning architecture at the session level to model DeepMood~\cite{cao2017deepmood}, a deep learning architecture based on late fusion.
However, in real life, the amount of data held by different medical institutions is very different, the number of medical institutions in different regions is also different. 
To this end, we divided our work into two parts. 
Firstly, we distribute the data to different parties for training according to the IID(Independent and Identically Distributed) method, but the amount of data that each party has during each training is not the same, and the number of parties participating in the training is also different each time. 
This setting simulates the real situation in different regions and different medical institutions. 
At the same time, in order to verify the influence of federated learning on model training, we simulated a data island environment and set up local training for each party that did not participate in federated learning. 
Furthermore, we assign the data to each party according to non-IID, and discuss the influence of non-IID on the prediction results. 
The experimental results show that the model prediction accuracy reaches 85.13\% in the case of IID and 76.95\% in the case of non-IID.

The rest of this article is organized as follows. 
The section~\ref{sec:preliminary} section introduces the background of multi-perspective learning, federated learning and privacy protection. At the same time, we analyzed the principle of the later fusion model. 
The task definition and the federated learning framework are described in Section~\ref{sec:method}. 
The data sources, experimental settings and results are outlined section~\ref{sec:exp}. 
Finally, we summarize the paper in section~\ref{sec:conclu}.

\begin{figure*}[!ht]
	\centering	
	\subfigure[Fully connected layer.]{
		\includegraphics[width=0.25\textwidth]{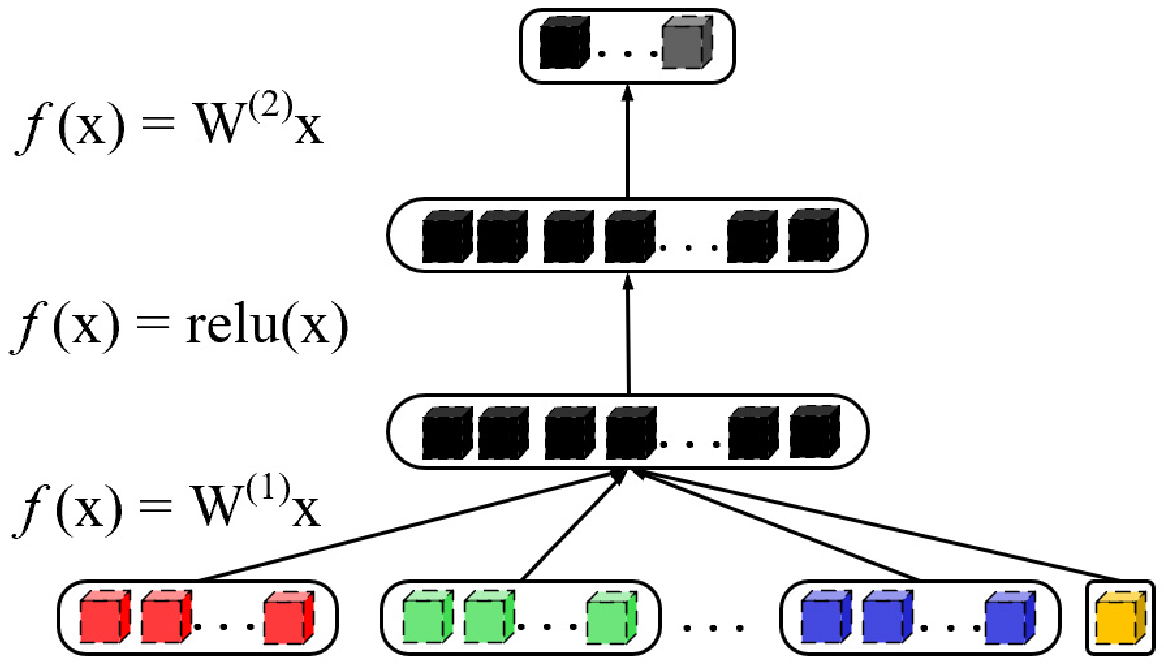}
		\setlength{\leftskip}{-20pt}
		\label{fig:back1}
	}
	\subfigure[Factorization Machine layer.]{
		\includegraphics[width=0.29\textwidth]{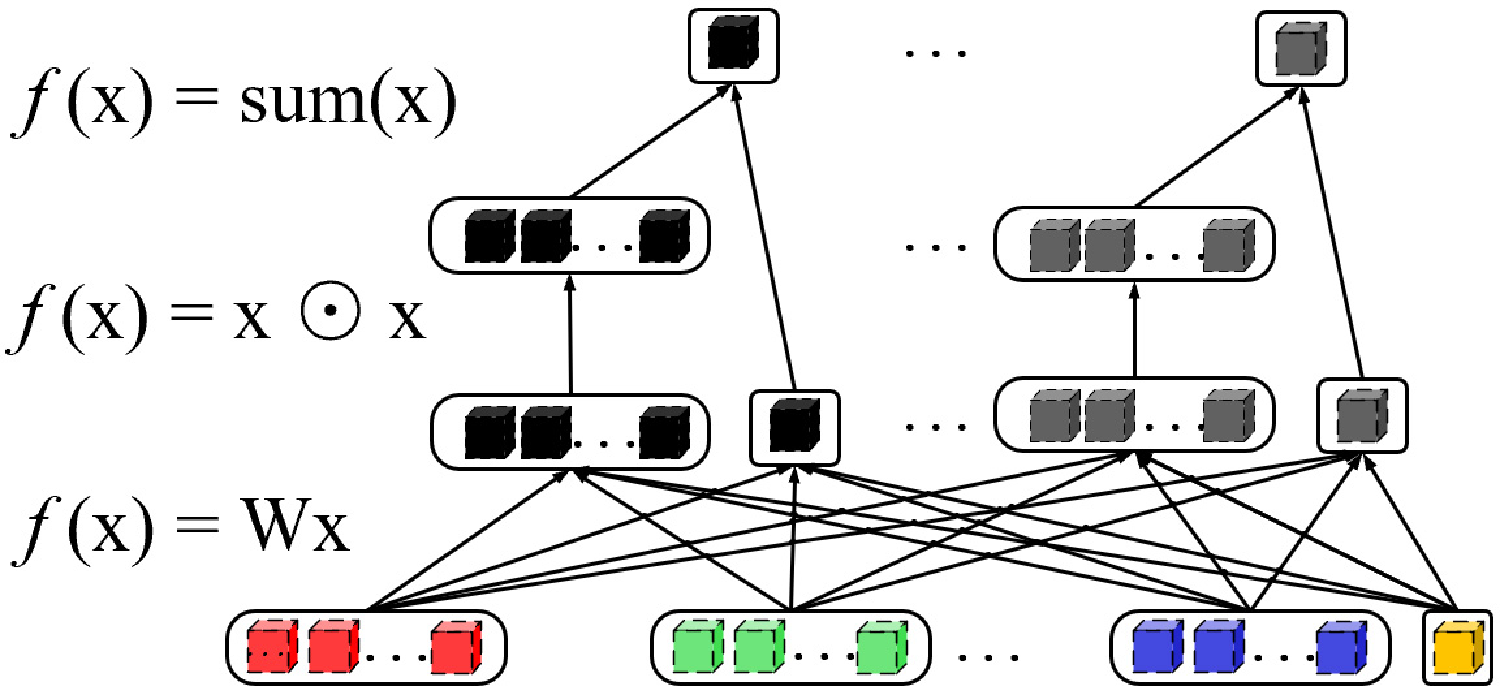}
		\setlength{\leftskip}{-20pt}
		\label{fig:back2}
	}
	\subfigure[Multi-view Machine layer.]{
		\includegraphics[width=0.36\textwidth]{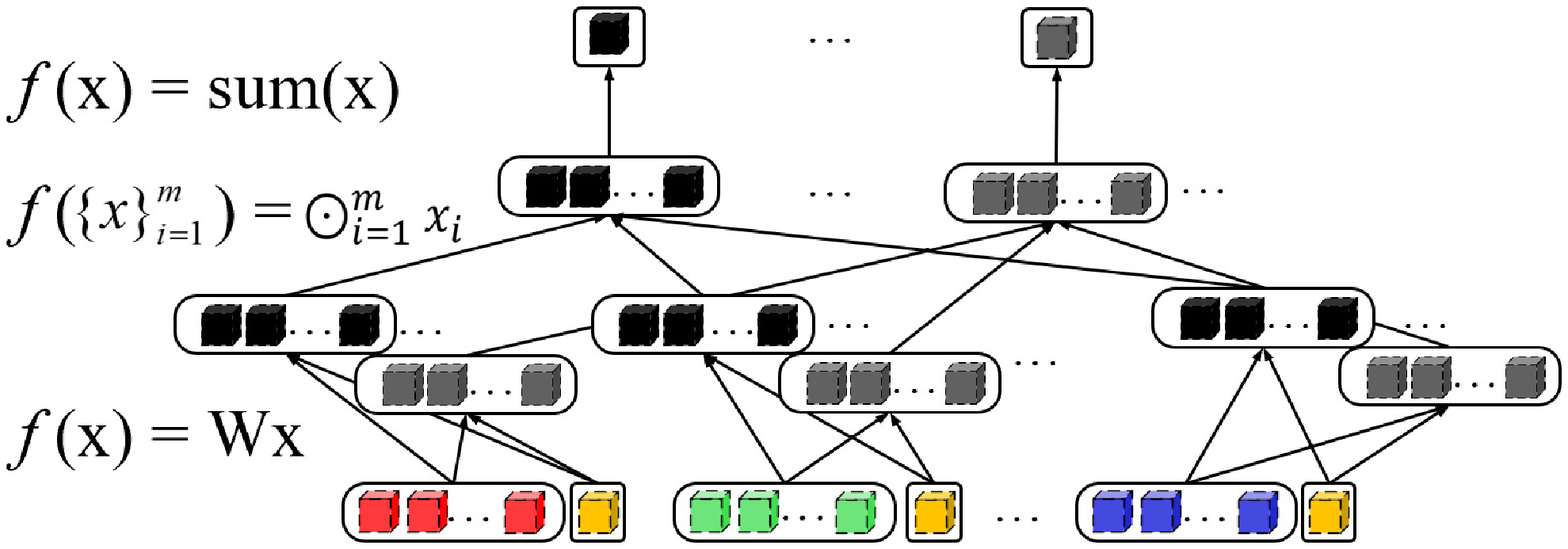}
		\setlength{\leftskip}{-20pt}
		\label{fig:back3}
	}
	\vspace{-1em}
	\caption{A comparison of different strategies for fusing multi-view data from the perspective of computational graph \cite{cao2017deepmood}.}\label{fig:back}
	\vspace{-1em}
\end{figure*}

\section{Background}\label{sec:preliminary}
\subsection{Related Work}
In this section, we introduce the related research results of 
federated learning and multi-view learning, and discuss the recently proposed federated multi-view learning.

\noindent\textbf{Multi-View learning.} Xu et al. \cite{xu2013survey} pointed out that multi-perspective learning requires the use of one function to model one perspective and the use of other perspectives to jointly optimize all functions. Cao et al. \cite{cao2016multi} used tensor product to process multi-view data. 
Yao et al. \cite{yao2018deep} integrated CNN, LSTM and Graph embedding tackled the complex nonlinear spatial and temporal dependency in a multi-perspective way. In addition, some work integrates multi-view into the process of deep learning \cite{zhang2016identifying} and transfer learning \cite{wu2017online}, so as to help expand samples from data.

\noindent\textbf{Federated learning.} Here we mainly refer to the medical application of federated learning and some common federated multi-View deep learning framework.

Lee et al. \cite{lee2018privacy} proposed a privacy protection platform in the federated environment, which could find similar patients from different hospitals without sharing patient-level information. 
Huang L et al. \cite{huang2019patient} improved the performance of federated learning for predicting mortality and length of stay by using feature autoencoders and patient clustering. 
However, the method of combining federated learning and multi-view data in existing research is still in the development stage.
Adrian Flanagan et al. \cite{flanagan2020federated} proposed the federated multi-view matrix factorization method and address cold-start problem.
Huang et al. \cite{huang2020federated} proposed FL-MV-DSSM, which is the first general content-based joint multi-view framework, which successfully extended traditional federated learning to federated multi-view learning.
Kang et al. \cite{kang2020fedmvt} proposed the FedMVT algorithm for semi-supervised learning, which can improve the performance of vertical federated learning with limited overlapping samples.
These related methods using public datasets are mainly used to recommend systems to solve the cold start problem. 
However, our framework uses data collected from mobile devices to solve medical mood prediction problems.

\noindent\textbf{Federated privacy protection.}
There are three main ways \cite{lyu2020privacy} to protect data privacy in the federated learning framework: Secure Multi-Party Computing (SMC) and Differential Privacy Mechanism (DP) and Homomorphic Encryption (HE).

Secure multi-party computing mainly uses secure communication and encryption algorithms to protect the model aggregation security of different participants in the federated learning \cite{canetti1996adaptively}. Since the federated framework does not need to aggregate data but transmits gradients or model parameters, SMC only needs to encrypt related parameters, which saves a lot of encryption calculation cost. However, the improved strategy based on SMC still adds extra time cost compared with the traditional federated framework. How to balance the time cost and the loss of data value over time has become a problem to be solved.

Differential privacy protects data privacy by adding noise to the data source, while ensuring that the loss of data quality is controllable \cite{Dwork2011}. By adding noise to the models or gradients uploaded by participants, the contribution of personal data in the dataset is masked to prevent reverse data leakage.
At the same time, because of the problem that the data after adding noise is still close to the original data, Sun L et al. \cite{sun2020federated,sun2020ldp} use local differential privacy and noise-free differential privacy mechanisms to decrease the risk of information exposure.
 However, the introduction of differential privacy may reduce the accuracy of the global model, and it will be difficult for the central server to measure the contribution of each party to encourage different parties to participate in the federated.

Homomorphic encryption can calculate the ciphertext data without decryption \cite{acar2018survey}. In the federated framework, each party can homomorphically encrypt the parameters they want to upload, and the central server can complete the aggregation process of the federated model without decryption. Since data and models will not be transmitted in plain text, there will be no leakage of the original data level. However, local party encryption and decryption operations increase computing power consumption, and the transmission of ciphertext will also increase additional communication cost.

\subsection{Later Fusion Model}
Since the datasets we use has the problem that the time series under three views have different frequencies and cannot be aligned, in this section, we introduce the later fusion strategy adopted by the model to make the time series of the data consistent \cite{rendle2012factorization,cao2016multi}. 
We set the output vector at the end of the $p-th$ view sequence as $k^{(p)}$, and let $\{k^{(p)}\in R^{d_{k}}\}_{p=1}^{n}$ be the multi-view data where $m$ is the number of views.

\noindent\textbf{Fully connected layer.}
We first consider the simplest way to connect multi-views directly, ie, $k=[k^{(1)};k^{(2)};...;k^{(n)}] \in R^{d}$, where $d$ is the total number of multi-view features, and typically $d=(2)nd_{k}$ for one directional(bidirectional) GRU. 
The connected hidden state $k$ is inserted into the fully connected neural network through a nonlinear function $\sigma(\cdot)$.
The feature interaction mode of the input unit is as follows:

\begin{equation}
\begin{split}
&p=relu(W^{(1)}[k;1]),\\
&\hat{y}=W^{(2)}p,
\label{eq:01}\vspace*{-10pt}
\end{split}
\end{equation}
where $W^{(1)}\in R^{k\times (d+1)}$ , $W^{(2)}\in R^{c\times h}$ , $h$ is the number of hidden units, $c$ is the number of classes, and the constant signal “1” is to model the global bias.
To simplify the illustration, we only set a hidden layer as shown in Fig~\ref{fig:back1}.

\noindent\textbf{Factorization Machine layer.}
As shown in Fig~\ref{fig:back2}, instead of transforming the input with a nonlinear function, we directly model the features of each input part as follows:

\begin{equation}
\begin{split}
&p_{c}=U_{c}k,\\
&b_{c}=W^{T}_{c}[k;1],\\
&\hat{y_{c}}=sum([p_{c}\odot{p_{c}};b_{c}]),
\label{eq:02}\vspace*{-10pt}
\end{split}
\end{equation}
where $U_{c}\in R^{f\times d}$ , $W_{c}\in R^{d+1}$, $f$ is the number of factor units, and $c$ denotes the $c$-th class.

\noindent\textbf{Multi-view Machine layer.}
Only considering the second-order feature interaction of the input data may not be comprehensive enough. 
We nest interaction to the $m$-th order between $m$ views to generate the final output $\hat{y_{c}}$  for the $c$-th class in the following way:

\begin{equation}
\small
\hat{y_{c}}=\beta_{0}+\sum_{v=1}^{m}\sum_{i_{v}=1}^{d_{v}}\beta_{i_{v}}^{(v)}k_{i_{v}}^{(v)}+\cdots+\hfill
\sum_{i_{1}=1}^{d_{1}}\cdots\sum_{i_{m}=1}^{d_{m}}\beta_{i}(\prod_{p=1}^{m}k_{i_{v}}^{(v)})\hfill,
\label{eq:add1}
\end{equation}
where $\beta$ is the global offset, the second part is the first-order fusion, and the last part is the $m$-th order fusion.
Next, the output vector $k_{i_{v}}^{(v)}$ is combined with the constant 1 as an additional feature. The Eq.~\ref{eq:add1} can be rewritten as follows:

\begin{equation}
\hat{y_{c}}=\sum_{i_{1}=1}^{d_{1}+1}\cdots\sum_{i_{m}=1}^{d_{m}+1}\omega_{i_{1},\cdots,i_{m}}(\prod_{v=1}^{m}[k_{i_{v}}^{(v)}:1]),
\label{eq:add2}
\end{equation}
where $\omega_{d_{1}+1,\dots,d_{m}+1}=\beta_{0}$ and $\omega_{i_{1},\dots,i_{m}}=\beta_{i_{1},\dots,i_{m}}$,  $\forall i_{v}\leq d_{v}$. Next, we decompose the $m$-th order weight tensor $\omega_{i_{1},\dots,i_{m}}$ into $k$ factors: $C\times U^{(1)}\times \cdots \times U^{(m)}$. $U^{(m)}\in R^{k\times(d_{h}+1)}$ is the factor matrix of the $m$-th view and $C \in R^{k\times \cdots \times k}$ is the identity tensor. Finally, we transform Eq.~\ref{eq:add2} as follows:

\begin{equation}
\hat{y_{c}}=\sum_{i_{1}=1}^{d_{k}+1}\cdots\sum_{i_{m}=1}^{d_{k}+1}(\sum_{f=1}^{h}\prod_{v=1}^{m}[k_{i_{v}}^{(v)}:1](i_{v})).
\label{eq:add3}
\end{equation}
As shown in the figure~\ref{fig:back3}, we can simplify Eq.~\ref{eq:add3} as follows:
\begin{equation}
\begin{split}
&p^{(v)}_{c}=U^{(v)}_{c}[k^{(v)};1],\\
&\hat{y_{c}}=sum([p^{(1)}_{c}\odot...\odot{p^{(m)}_{c}}]).
\label{eq:03}\vspace*{-10pt}
\end{split}
\end{equation}

\noindent where $U_{c}^{v} \in R^{h\times d_{k}+1}$
is the factor matrix of the $v$-th view for
the $c$-th class.$\hat{y_{c}}$ is the final output for the $c$-th class. 

\section{Iot-data silo island problem and methodology}~\label{sec:method}
In this section, we introduce how to use Iot-data to train local and federated learning models. 
We first discuss the reasons for task definition, and then introduce the federated learning framework proposed by Google \cite{mcmahan2016federated}.

\subsection{Problem Description}
In the absence of federated learning frameworks, medical institutions can only use local datasets without interactive processes when using machine learning algorithms to build models for disease diagnosis, medical imaging research, and so on.
We retain the local learning model as a comparison to measure the improvement effect of the federated learning algorithm on the multi-views heterogeneous data training model. 
We have conceived the following three situations.

At first there were several hospitals in a city $\{H_{1},...,H_{m}\}$.
Assume that patients with bipolar I disorder and bipolar II disorder and normal people who are suspected of being sick will go to different hospitals for depression test scores on average, and the hospital will also record the patient's mobile terminal data. 
From a certain moment, we stop the collection of data by hospitals. 
At this time, each hospital has a fixed amount of data $D_{x}$. 
Each medical institution will first use its own local data to train and test the effect. 
The results obtained at this time are generally difficult to use as a reference for the diagnosis of depression. 
Each participant will cooperate with other medical institutions for federated study and training, and in this process, new medical institutions will continue to participate. 
Without reducing the total number of communication rounds, increase the degree of parallelism to test the changes in the prediction effect.

At a certain moment, the number of hospitals in a certain city is constant at $n$ $\{H_{1},...,H_{n}\}$, and no new hospitals will be established in this city for a certain period of time. 
Patients will go to different hospitals on average as described above, and all medical institutions will predict depression mood through local training and federated learning. 
Initially, each hospital has a small amount of data $D_{a}$. 
As patients continue to come to the hospital for treatment and review, the hospital will continue to increase the amount of data $D_{x}$. 
When the data added by each hospital reaches a mark value $D_{f}$, the participants will restart the training in hopes of improving the prediction effect.

In the actual medical environment, the data owned by each hospital must be non-IID. 
We assume that patients with bipolar I disorder and bipolar II disorder and normal people who are suspected of being sick will only go to a specific hospital $H_{x}$ for treatment.
Each medical institution has a different amount of patient data, and the serious condition of patients is inconsistent, resulting in extreme data distribution. 
On this basis, we detect the impact of this extreme distribution data on the decline of model prediction progress compared to IID data. 
The specific data division is introduced in section~\ref{sec:compare}.

\begin{figure}[!tpb]
\centering
\includegraphics[width = 9 cm]{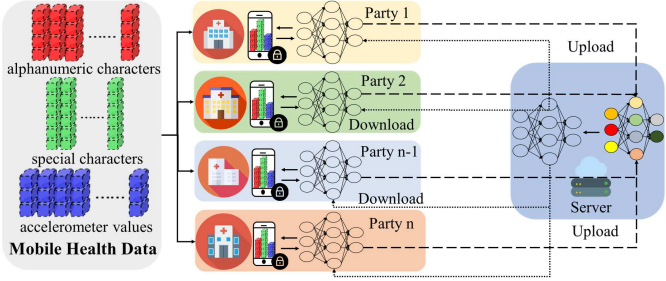}
\caption{The architecture of federated learning.
Firstly, the participated parties have data on normal people, bipolar I and bipolar II users, and no data interaction between different parties. 
At the beginning of each communication round, the server will assign the global model to the participated party in training in this round.
Next, the activated parties will train the local model through its own mobile health data and upload it to the server. 
Finally, the server updates the global model according to the uploaded local model.}\label{fig:02}
\end{figure}

\subsection{Federated learning}
Due to the privacy issues of patient health data stored in hospitals, we cannot use these data for centralized learning. 
Google has proposed a new framework called federated learning. 
During the training of the federated learning model, the data owned by each hospital participating in the model training collaboration can be saved locally without uploading. 
Each hospital uses its own data to download the model from the server for training, and upload the trained model or gradient to the server for aggregation, and then the server sends the aggregated model or gradient information to each hospital. 
Considering the communication burden and connection reliability and other issues, we adopt the model average method for training.
Assuming that there are $K$ hospitals participating in federated learning, when the global model parameters are updated in the $t$ round, the $K$-th participant calculates the local data average gradient of the current model parameters according to Eq.~\ref{eq:04}, and the server aggregates these gradients and uses these model parameters to update the global model according to Eq.~\ref{eq:05}.

\begin{equation}
g_{k}=\nabla F_{k}(\omega_{t}),
\label{eq:04}
\end{equation}

\begin{equation}
\omega_{t+1}\leftarrow \omega_{t}-\eta \sum^{K}_{k=1}\frac{n_{k}}{n}g_{k},
\label{eq:05}
\end{equation}
\noindent where $g_{k}$ is the average gradient of the local data of the current model parameter $\omega_{t}$. $\eta$ is the learning rate, $\sum^{K}_{k=1}\frac{n_{k}}{n}g_{k}=\nabla f(\omega_{t})$ .
Literature \cite{mcmahan2016federated} also proposed an equivalent federated model training method.

Each hospital uses local data to perform one (or more) steps of gradient descent, according to Eq.~\ref{eq:06} the existing model parameters locally and sends the locally updated model parameters to the server. 
The server then calculates the weighted average of the model results according to Eq.~\ref{eq:07} and sends the aggregated model parameters to each hospital.

The literature \cite{mcmahan2016federated} shows that compared with the purely distributed SGD, the improved scheme can reduce the amount of communication used by 10-100 times, and can choose the update optimizer of the gradient other than SGD.

\begin{equation}
\forall k,\omega^{(k)}_{t+1}\leftarrow \overline{\omega}_{t}-\eta g_{k},
\label{eq:06}
\end{equation}

\begin{equation}
\overline{\omega}_{t+1}\leftarrow \sum^{K}_{k=1}\omega^{(k)}_{t+1},
\label{eq:07}
\end{equation}
\noindent where $\overline{\omega_{t}}$ is the existing model parameter of the local client.

In this work, we use the federated learning framework to focus on studying the influence of participants and data volume under IID data and the extreme distribution of non-IID data on the results of emotion prediction through multi-views heterogeneous data collected on the mobile terminal. 
However, in the actual use of the federated learning framework, there are also such issues as the flexibility and stability of party joining and quitting at any time, the party's dynamic increase and change with the increase of patients, the statistical contribution of party to the model and the design of corresponding incentive mechanism.
Although these issues are beyond the scope of our existing work, we still deal with non-IID data and party participation stability. 
We specified the number of parties participating in the experiment. 
The datasets owned by each client have been determined before training, and we fixed the parties participating in the training update in advance to stabilize the final training effect after each experiment.

\begin{algorithm}[t]  
  \caption{Federated Averaging. The $K$ is the total number of party, $M$ is the local minibatch size, $E$ is the number of local epochs, and $\eta$ is the learning rate.}  
    \textbf{Server executes:}\\
        \quad Initialize $\omega_{0}$\\
        \quad\textbf{for} each round $t=1,2,…$ \textbf{do}\\
            \qquad$t\leftarrow$ random choose$(1,K)$\\
            \qquad$C_{t}\leftarrow$(random set of $t$ parties)\\
            \qquad \textbf{for} each party $k\in C_{t}$ in parallel \textbf{do}\\
                \quad \qquad $\omega_{t+1}^{(k)}\leftarrow localtraining(k,\overline{\omega}_{t})$\\
            \qquad$\overline{\omega}_{t+1}\leftarrow \sum^{K}_{k=1}\frac{n_{k}}{n}\omega_{t+1}^{(k)}$\\
    \quad\\
    \textbf{Localtraining}$(k,\overline{\omega}_{t})$ \textbf{:} $//k$ parties training in parallel\\
        \quad\textbf{for} each local epoch $i$ from $1$ to $S$ \textbf{do}\\
            \qquad $D_{m}\leftarrow$(split $D_{k}$ into batches of size $M$ randomly)\\
            \qquad \textbf{for} each batches $b$ from $1$ to $B=\frac{n_{k}}{M}$ \textbf{do}\\
                \quad\qquad $\omega_{b+1}^{(k)}\leftarrow\omega_{b,i}^{(k)}-\eta\nabla F_{k}(\omega_{t})$\\
        \quad return $\omega_{t+1}^{(k)}=\omega_{B,S}^{(k)}$ to server\\
    \label{code:recentEnd}  
\end{algorithm}

\section{Experiments}~\label{sec:exp}
In this section, we introduce how to use data generated by personal mobile devices to train deep learning models. 
We assume that the hospital allocates a special mobile device to each user to collect alphanumeric characters, special characters, and accelerometer values used in the conversation, and the hospital has weekly HDRS test scores for patients. 
Due to the particularity of the diagnosis of depression, patients may go to multiple hospitals to try and seek treatment, and some hospitals may have the same patient data.

\subsection{Dataset}

The data used in the experiment comes from a real observation study of a free mobile app by BiAffect. 
In the data collection stage, the researchers provide the users with a special Android smartphone. 
The mobile phone uses a customized virtual keyboard to replace the default keyboard, so as to collect the metadata input by the user without affecting the operation in the background. 
The collected content of the keyboard includes data such as the user key input time, the number of keystrokes and the phone accelerometer value. 
The three types of metadata we use are as follows:

\noindent\textbf{Alphanumeric characters.} In order to protect user privacy, we do not collect specific alphanumeric characters. 
We only collect the duration and time of the button, the duration after the last button was pressed, and the distance from the last button to the coordinate axis on the horizontal and vertical axes.

\noindent\textbf{Special characters.} We perform one-hot encoding for operations including space, backspace, and keyboard switching. 
Compared with alphanumeric characters, the number of operations for special characters is less.

\noindent\textbf{Accelerometer value.} The accelerometer record every 60ms between sessions. 
Because different users have different typing speeds, the accelerometer values are more densely recorded than alphanumeric characters.

We define a session as an interval of more than five seconds after the user presses the key five seconds after the last key press or lasts longer. 
Due to the user typing habits, the duration of the session is generally less than one minute.

At the same time, participants receive the Hamilton Depression Rating Scale (HDRS) \cite{hamilton1986hamilton} and the Young Man Mania Scale (YMRS) \cite {young1978rating} once a week, which is a doctor clinical diagnosis and evaluation of bipolar depression and a very effective assessment questionnaire for bipolar disorder. 
After the data collected by the study participants, according to the control of extreme subjects and normal users. 
There are 6 participants suffered from bipolar I disorder, including severe episodes ranging from bipolar disorder to depression, 6 participants suffered from bipolar II disorder, including clinical manifestations were mildly elevated mood between mild manic episodes and severe episodes, and 8 participants were diagnosed as normal subjects.
Since the evaluation process only relies on the communication between the patient and the doctor and the indicators given by the evaluation scale, the results of the diagnosis are not necessarily reliable 
Therefore, we try to predict the occurrence of depression from an objective perspective by recording real-time data of patients.

\subsection{Experimental Setup}

Our model is implemented using Keras with Tensorflow as the backend.
All experiments are conducted on a 64 core Intel Xeon CPU E5-2680 v4@2.40GHz with 512GB RAM and 1$\times$ NVIDIA Tesla P100-PICE GPU.
We use RMSProp \cite{tieleman2012lecture} as the training optimizer.
We retain sessions with keypresses between 10 and 100, and finally generate 14960 samples. 
Each user contributes first 80\% sessions for training and the rest for validation.

\begin{table}[h]
\caption{Parameter configuration.}
\centering
\label{tab:02}
\begin{tabular}{c|cc}
\toprule  
Parameter& \qquad\qquad\qquad &Value \\
\midrule  
DNN communication rounds&\qquad &400 \\
DNN local epochs&\qquad &15 \\
DFM communication rounds&\qquad &300 \\
DFM local epochs&\qquad &20 \\
DMVM communication rounds&\qquad &400 \\
DMVM local epochs&\qquad &15 \\
Batch size&\qquad &256 \\
Learning rate&\qquad &0.001 \\
Dropout fraction&\qquad &0.1 \\
Maximum sequence length&\qquad &100 \\
Minimum sequence length&\qquad &10\\
\bottomrule 
\end{tabular}
\end{table}

\begin{table}[h]
\caption{The accuracy of the compared models under different local epochs and communication rounds. We show the best results with boldface.}
\label{tab:01}
\begin{tabular}{c|c|ccccc}
\hline
\multicolumn{2}{c|}{Communication Rounds} & \multirow{2}{*}{100} & \multirow{2}{*}{200} & \multirow{2}{*}{300} & \multirow{2}{*}{400} & \multirow{2}{*}{500} \\ \cline{1-2}
Model                 & local epochs &        &                 &                 &                 &                 \\ \hline
\multirow{4}{*}{DNN}  & 5            & 0.7919 & 0.8295          & 0.8435          & \textbf{0.8501} & 0.8442          \\
                      & 10           & 0.8005 & 0.8358          & 0.8498          & \textbf{0.8525} & 0.8468          \\
                      & 15           & 0.8335 & 0.8618          & 0.8635          & \textbf{0.8638} & 0.8521          \\
                      & 20           & 0.8472 & 0.8472          & \textbf{0.8521} & 0.8388          & 0.8372          \\ \hline
\multirow{4}{*}{DFM}  & 5            & 0.8188 & 0.8435          & \textbf{0.8491} & 0.8481          & 0.8462          \\
                      & 10           & 0.8182 & 0.8432          & 0.8448          & 0.8448          & \textbf{0.8465} \\
                      & 15           & 0.8338 & \textbf{0.8518} & 0.8468          & 0.8448          & 0.8388          \\
                      & 20           & 0.8402 & 0.8525          & \textbf{0.8531} & 0.8501          & 0.8415          \\ \hline
\multirow{4}{*}{DMVM} & 5            & 0.7892 & 0.8425          & 0.8531          & \textbf{0.8685} & 0.8518          \\
                      & 10           & 0.8139 & 0.8468          & \textbf{0.8601} & 0.8578          & 0.8472          \\
                      & 15           & 0.8261 & 0.8448          & 0.8568          & \textbf{0.8695} & 0.8395          \\
                      & 20           & 0.8101 & 0.8198          & \textbf{0.8288} & 0.8245          & 0.8255          \\ \hline
\end{tabular}%
\end{table}

\begin{figure*}[t]
	\centering	
	\subfigure[Alphanumeric characters.]{
		\includegraphics[width=0.3\textwidth]{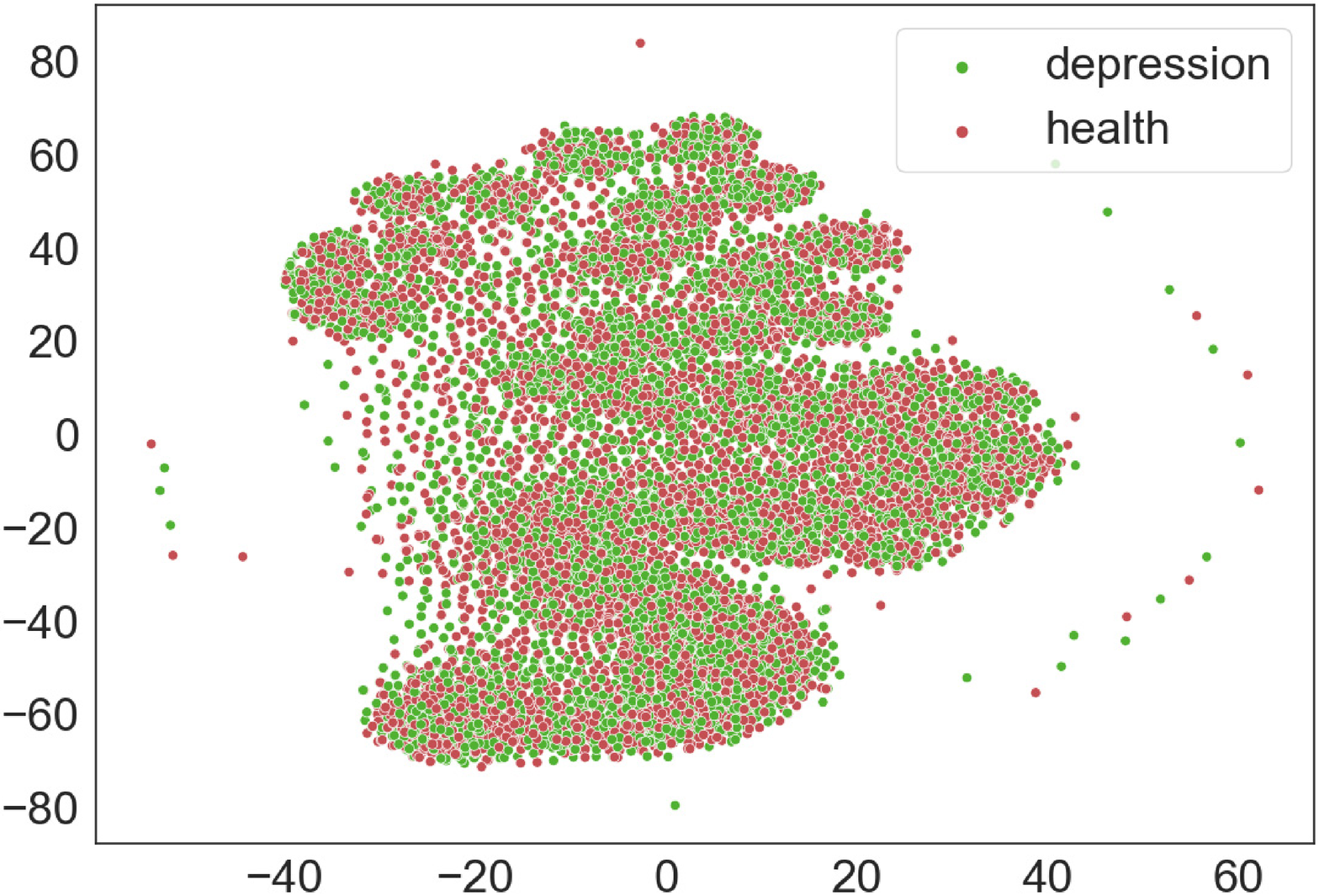}
		\label{fig:predis_svr}
	}\ \ \
	\subfigure[Special characters.]{
		\includegraphics[width=0.3\textwidth]{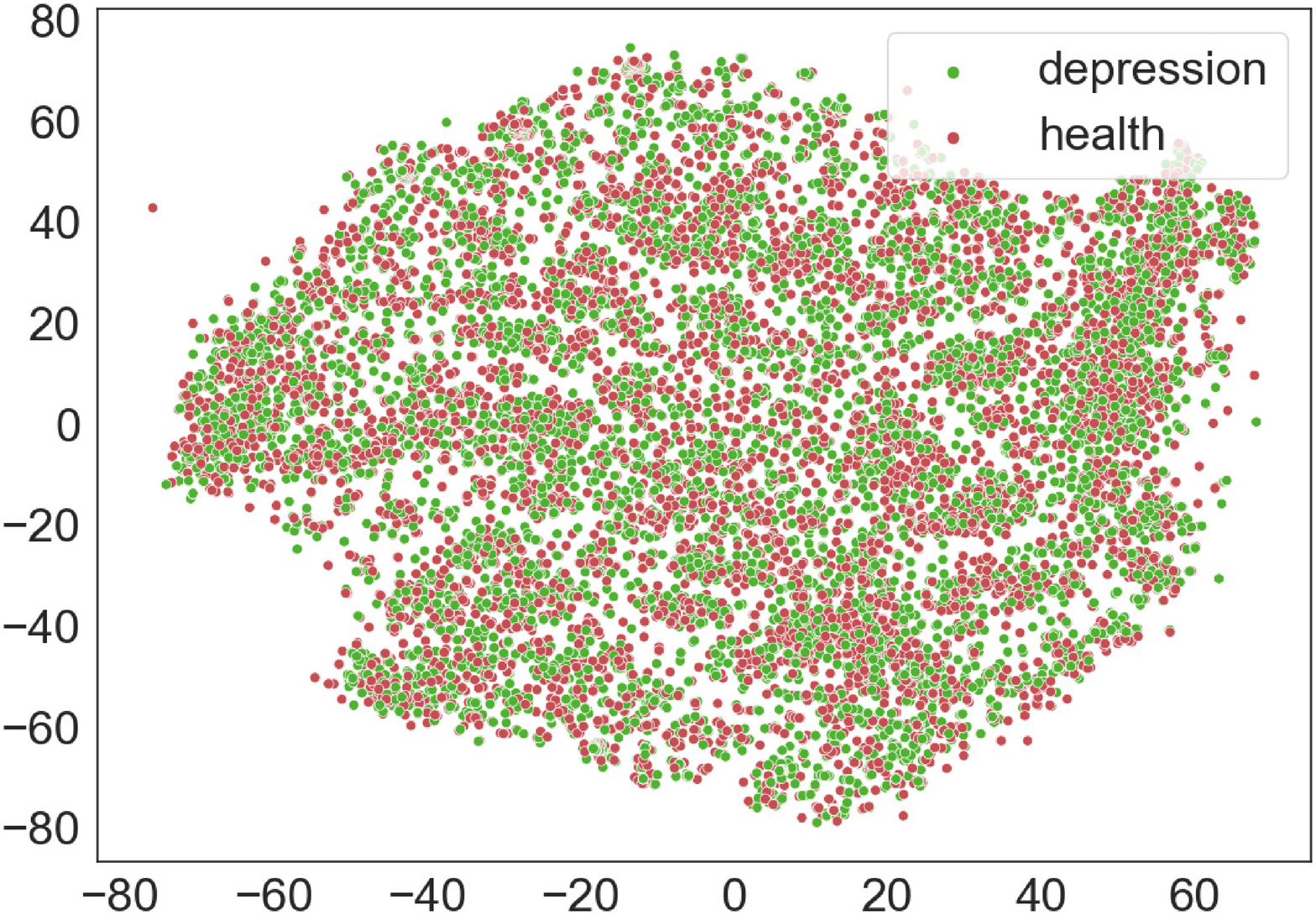}
		\label{fig:predis_lstmd}
	}\ \ \
	\subfigure[Accelerometer value.]{
		\includegraphics[width=0.3\textwidth]{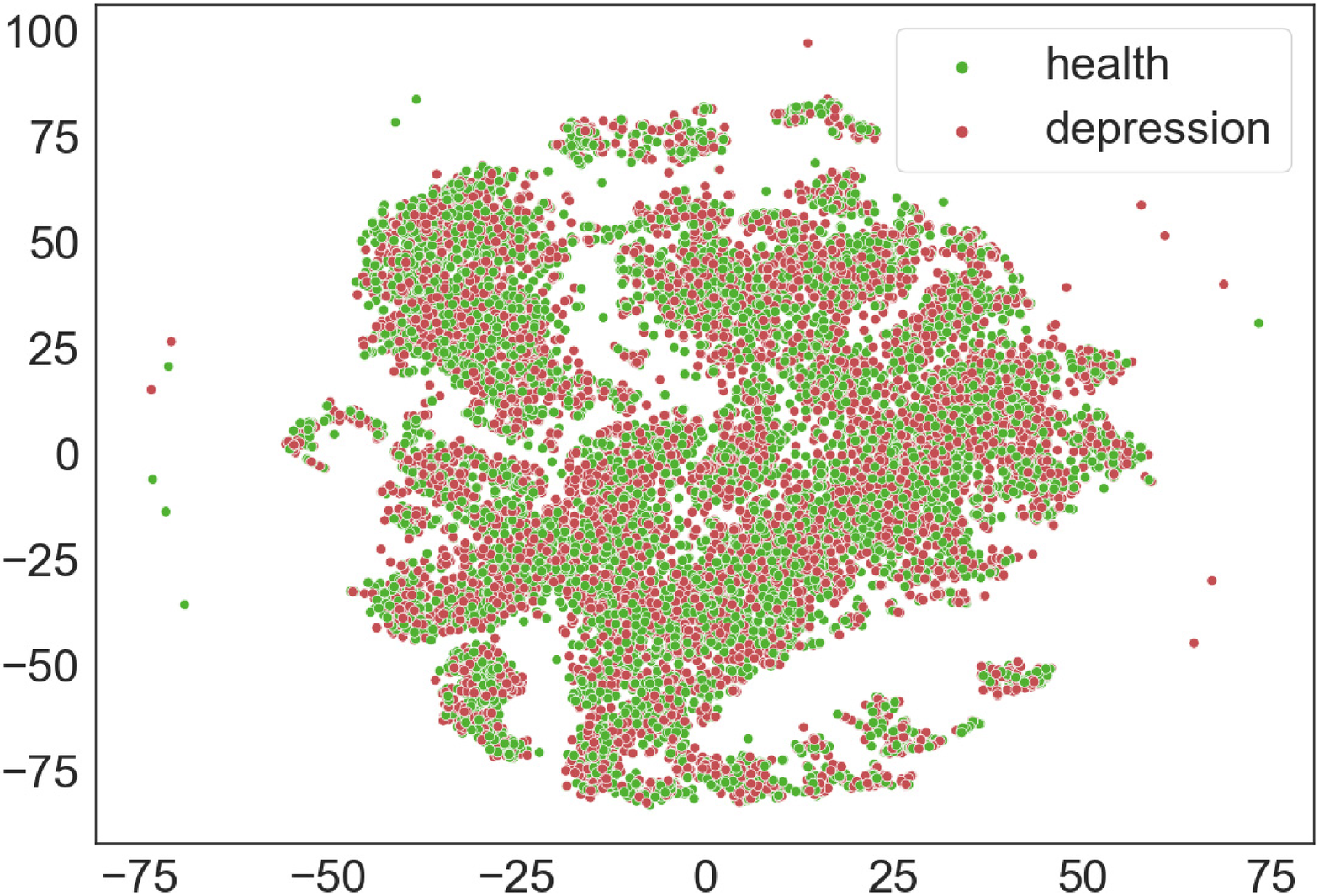}
		\label{fig:predis_gcn}
	}
	\vspace{-1em}
	\caption{Visualization of labeling with TSNE for three views}\label{fig:03}
	\vspace{-1em}
\end{figure*}

We set the parameters based on experience and some experimental comparisons, including the number of communication rounds, the number of local epochs, batch size, learning rate, and dropout rate. 
We consider sessions with the HDRS score between 0 and 7 (inclusive) as negative samples (normal) and those with HDRS greater than or equal to 8 as positive samples (from mild to severe depression). 
Our code is open-sourced at \url{https://github.com/RingBDStack/Fed_mood}.

In order to study the influence of local epochs parameters, we evenly distribute the training data set to 8 participants for testing.
The results are shown in the Table~\ref {tab:01}, we can find: (1) As the number of communication rounds increases, the accuracy rate shows a trend of rising and then a slight decrease.
(2) Our work is inconsistent with the results of Zhao et al. \cite{zhao2018federated}. 
A large number of local epochs can significantly improve the effect of federated learning. 
However, when epochs = 20, the accuracy of DMVM in any communication round shows a downward trend.
These results show that increasing the local epoch can make the training more stable and speed up the convergence speed, but it may not make the global model converge to a higher accuracy level. 
That is, over-optimizing the local datasets may cause performance loss. 
(3) In the first 300 periods, the fusion efficiency of DFM is higher than that of DNN and DMVM, which shows the improvement effect of the fusion layer, and DFM achieve better local minima of loss functions in some results.
Compared with centralized learning, due to the sharp reduction of local data, the effect of DMVM fusion of multi-view and multi-level features will be affected to a certain extent.
Because the three models get different results when the local epochs are 15 and 20, we perform the parameters separately settings, as shown in Table~\ref{tab:02}.

\subsection{IID Experiments}\label{sec:compare}
\subsubsection{Compared Methods}
We compare FedAVG with the following methods, each of which represents a different strategy for data interaction.

\noindent\textbf{Local Training}: Local training means that each party only uses its data for training, without any interaction with other parties.

\noindent\textbf{CDS \cite{sheller2020federated}}: Collaborative data sharing is a traditional centralized machine learning strategy, which requires that each party uploads its patient data to the center server for training.

\noindent\textbf{IIL \cite{sheller2020federated,chang2018distributed}}: Institutional incremental learning is a serial training method. Each party transfers its model to the next participator after training finish, until all have trained once.

\noindent\textbf{CIIL \cite{sheller2020federated,chang2018distributed}}: Cyclic institutional incremental learning repeats the IIL training process. It keeps consistent with the number of federated learning local training epochs and looping repeatedly through the parties.

In each experiment, the models we compared are summarized as follows:

\noindent\textbf{DMVM}: The proposed DeepMood architecture with a Multi-view Machine layer for data fusion.

\noindent\textbf{DFM}: The proposed DeepMood architecture with a Factorization Machine layer for data fusion.

\noindent\textbf{DNN}: The proposed DeepMood architecture with a conventional fully connected layer for data fusion.

In this work, for the IID setting, we randomly assign each client a uniform distribution of three data categories: normal users, bipolar I disorder patients, and bipolar II disorder patients.The specific methods as follows:

1.The number of participants data remains unchanged, and the number of parallel participants is increasing. The amount of data owned by each party is fixed at 1500, and the number of hospitals participating in the training gradually increases from 4. We test the training effect of up to 24 parallel participants.

2.The number of concurrent participants remains the same, increasing the amount of data owned by each participant. Consistent with the experiment of setting hyperparameters, we set the number of concurrent participants to 8. The amount of data owned by each party gradually increases from 100, and we use about 25\% (3000) of the total data as the maximum value of the experiment.

In order to make the results of the experiment stable, we conduct each group of experiments five times and average the results.

\subsubsection{Evaluation criteria}
In order to evaluate the influence of federated learning and local training on the prediction results, we adopted the following measures:

Accuracy is one of the most frequently used criterion, which represents the ratio of the number of correctly predicted samples to the total number of predicted samples.
In the federated learning experiment, the central server can test the final global model with its own test set. 
In the local training experiment, we regard the local data as a whole, and compare the number of samples correctly predicted by each participant with the test set.

\subsubsection{Experiment Result}
Table~\ref{tab:04} shows the mood prediction effect of increasing parallel parties.
Since local training has no interactive learning process and the amount of data owned by each participant is constant, the final result has always been between 73\% and 75\% fluctuation. 
In most cases, CDS can achieve the best prediction effect, but the best effect of DMVM model using CIIL can reach 85.29\%, which is about 18\% higher than local training without updating model weight.

Table~\ref{tab:05} shows the accuracy performance of increasing the amount of data for each participant. 
The accuracy of the local training without weight update and FedAvg training is increasing at the same time. 
When the amount of data for each party is small (data\textless1000), the improvement effect of FedAvg compared with local training can reach up to 16.7\%. 
When the amount of data for each participant is large enough (data=3000), the FedAvg enhancement effect is up to 10.5\%, which has a small difference from the result of CIIL.

\begin{table}[]
\centering
\caption{Accuracy performace of the IID experiments I. We show the best results with boldface.}
\label{tab:04}
\begin{tabular}{c|c|ccc}
\hline
Number of party     & Metrics        & DNN             & DFM             & DMVM            \\ \hline
\multirow{5}{*}{4}  & Local Training & 0.7431          & 0.7514          & 0.7343          \\
                    & CDS            & \textbf{0.8221} & 0.8144          & 0.7953          \\
                    & FedAVG         & 0.8114          & \textbf{0.8255} & 0.8095          \\
                    & IIL            & 0.7909          & 0.7851          & 0.7781          \\
                    & CIIL           & 0.8112          & 0.8107          & \textbf{0.8098} \\ \hline
\multirow{5}{*}{8}  & Local Training & 0.7344          & 0.7337          & 0.7267          \\
                    & CDS            & \textbf{0.8294} & \textbf{0.8366} & \textbf{0.8295} \\
                    & FedAVG         & 0.8283          & 0.8266          & 0.8156          \\
                    & IIL            & 0.7899          & 0.7774          & 0.7751          \\
                    & CIIL           & 0.8266          & 0.8288          & 0.8184          \\ \hline
\multirow{5}{*}{12} & Local Training & 0.7406          & 0.7444          & 0.7238          \\
                    & CDS            & 0.8352          & \textbf{0.8526} & 0.8309          \\
                    & FedAVG         & \textbf{0.8462} & 0.8304          & 0.8287          \\
                    & IIL            & 0.7971          & 0.7827          & 0.7703          \\
                    & CIIL           & 0.8324          & 0.8360          & \textbf{0.8409} \\ \hline
\multirow{5}{*}{16} & Local Training & 0.7356          & 0.7379          & 0.7248          \\
                    & CDS            & \textbf{0.8463} & \textbf{0.8507} & 0.8478          \\
                    & FedAVG         & 0.8383          & 0.8388          & 0.8281          \\
                    & IIL            & 0.7884          & 0.8039          & 0.7827          \\
                    & CIIL           & 0.8352          & 0.8358          & \textbf{0.8519} \\ \hline
\multirow{5}{*}{24} & Local Training & 0.7388          & 0.7455          & 0.7240          \\
                    & CDS            & \textbf{0.8642} & \textbf{0.8597} & 0.8479          \\
                    & FedAVG         & 0.8513          & 0.8429          & 0.8374          \\
                    & IIL            & 0.7945          & 0.8144          & 0.7940          \\
                    & CIIL           & 0.8493          & 0.8487          & \textbf{0.8529} \\ \hline
\end{tabular}
\end{table}

\begin{table}[]
\centering
\caption{Accuracy performace of the IID experiments II. We show the best results with boldface.}
\label{tab:05}
\begin{tabular}{c|c|lll}
\hline
Number of data & Metrics & \multicolumn{1}{c}{DNN} & \multicolumn{1}{c}{DFM} & \multicolumn{1}{c}{DMVM} \\ \hline
\multirow{5}{*}{100}  & Local Training & 0.6420          & 0.6230          & 0.6180          \\
                      & CDS            & \textbf{0.7439} & \textbf{0.7366}          & \textbf{0.7319}          \\
                      & FedAVG         & 0.7176          & 0.7172 & 0.7026          \\
                      & IIL            & 0.6295          & 0.6722          & 0.6126          \\
                      & CIIL           & 0.6958          & 0.7304          & 0.7043 \\ \hline
\multirow{5}{*}{500}  & Local Training & 0.6878          & 0.6894          & 0.6828          \\
                      & CDS            & \textbf{0.7982} & 0.7836 & \textbf{0.7811} \\
                      & FedAVG         & 0.7865         & 0.8044          & 0.7810          \\
                      & IIL            & 0.7389          & 0.7272          & 0.7244          \\
                      & CIIL           & 0.7612          & \textbf{0.8058}          & 0.7771          \\ \hline
\multirow{5}{*}{1000} & Local Training & 0.7192          & 0.7283          & 0.7129          \\
                      & CDS            & 0.8204          & 0.8098 & 0.8151          \\
                      & FedAVG         & \textbf{0.8309} & \textbf{0.8259}          & \textbf{0.8267}          \\
                      & IIL            & 0.7647          & 0.7707          & 0.7642          \\
                      & CIIL           & 0.8266          & 0.8183          & 0.8165 \\ \hline
\multirow{5}{*}{1500} & Local Training & 0.7267          & 0.7337          & 0.7344          \\
                      & CDS            & \textbf{0.8294} & \textbf{0.8366} & \textbf{0.8295}          \\
                      & FedAVG         & 0.8283          & 0.8266          & 0.8156          \\
                      & IIL            & 0.7899          & 0.7774          & 0.7751          \\
                      & CIIL           & 0.8279          & 0.8288          & 0.8184 \\ \hline
\multirow{5}{*}{2000} & Local Training & 0.7516          & 0.7558          & 0.7395          \\
                      & CDS            & 0.8331 & \textbf{0.8490} & 0.8349          \\
                      & FedAVG         & \textbf{0.8400}          & 0.8359          & 0.8303          \\
                      & IIL            & 0.8002          & 0.7920          & 0.7968          \\
                      & CIIL           & 0.8343          & 0.8470          & \textbf{0.8374} \\ \hline
\multirow{5}{*}{3000} & Local Training & 0.7726          & 0.7872          & 0.7591          \\
                      & CDS            & 0.8377 & \textbf{0.8612} & \textbf{0.8503} \\
                      & FedAVG         & 0.8432          & 0.8430          & 0.8390          \\
                      & IIL            & 0.8145          & 0.8130          & 0.8018          \\
                      & CIIL           & \textbf{0.8473}          & 0.8443          & 0.8501 \\ \hline
\end{tabular}
\end{table}

\begin{table}[]
\centering
\caption{Accuracy performance of Non-IID experiment and IID. We show the best results with boldface.}
\label{tab:03}
\begin{tabular}{c|c|ccc}
\hline
Types of data            & Metrics & DNN             & DFM             & DMVM            \\ \hline
\multirow{4}{*}{Non-IID} & CDS     & \textbf{0.8393} & \textbf{0.8228} & \textbf{0.8318} \\
                         & FedAVG  & 0.7695          & 0.7159          & 0.7684          \\
                         & IIL     & 0.6881          & 0.6881          & 0.7032          \\
                         & CIIL    & 0.7316          & 0.7416          & 0.7651          \\ \hline
\multirow{4}{*}{IID}     & CDS     & \textbf{0.8221} & 0.8144          & 0.7953          \\
                         & FedAVG  & 0.8114          & \textbf{0.8255} & 0.8095          \\
                         & IIL     & 0.7909          & 0.7851          & 0.7781          \\
                         & CIIL    & 0.8112          & 0.8107          & \textbf{0.8098} \\ \hline
\end{tabular}
\end{table}

\subsection{Non-IID Experiments}
In the real medical environment, the data owned by the hospital should be non-IID. In this section we introduce the methods and results of non-IID experiments.
\subsubsection{Compared Methods}
The models we compared are shown in section 4.3.1.
In this work, for non-IID settings, we have 8 normal users personal data, 6 bipolar I disorder patients data, 6 bipolar II disorder patients data. 
There are 4 hospitals participating in the training experiment, and the data volume of different users is inconsistent. 
Each hospital has two normal users data, one bipolar I disorder patients data, and one bipolar II disorder patients data.

\subsubsection{Evaluation criteria}
Our evaluation criteria are consistent with section 4.3.2, and accuracy is still used as the criterion for evaluating mood prediction.

\subsubsection{Experiment Result}
As shown in Table~\ref{tab:03}, the prediction effect of CDS under the non-IID setting is far ahead of the distributed cooperative learning method, and the federated learning prediction accuracy of the three models decreased by 5.2\% (DNN), 13.3\% (DFM) and 5.1\% (DMVM).  
The nature of the extreme distribution of non-IID data is the reason for the decline in prediction effect. 
We also find that the prediction effects of the two models that do not use nonlinear functions for feature interaction are significantly different under non-IID data. 
Due to the large difference in the number of patient data owned by each party and the completely inconsistent patient data types, the second-order feature interaction failed to integrate all features well, and the log also showed that its prediction accuracy fluctuates more than DMVM.

\subsection{Discussion}
In this section, we discuss the experimental results. 

As shown in Table~\ref{tab:04}, when the amount of data held by each party is constant, we can find that CDS can always maintain the best effect on DNN and DFM models in most cases, but in the DMVM model, CIIL has the best result. We can see from Table~\ref{tab:05} that when the amount of data is 1000, the data of each party is not repeated, and the federated learning framework has achieved the best results under the three models. When the amount of data is 1500, the model is affected by repeated data for the first time. Instead, the prediction performance of FedAVG declined slightly, and the prediction effect of CIIL is still rising.

Since the prediction accuracy of CIIL mostly depends on the effect of the last trained party model, we guess that repeated input data will seriously affect the fusion interaction mode of the multi-view machine layer. Compared with CDS, the last participation of CIIL has less repeated data, and the federated framework will be affected by repeated data due to the last round of global model weight update. Therefore, CIIL is less affected than these two methods, and the best accuracy result can eventually reach 85.29\%.

For non-IID experiments, CDS can still maintain the prediction accuracy of about 83\%, while federated learning and CIIL both have an inadequate accuracy drop. However, under the non-IID setting, the federated learning framework surpasses CIIL on the DMVM model, which proves the superiority of the multi-view machine layer under the federated framework.

At the same time, for the non-IID experiment, we also find that in one hospital, the accuracy of the validation set is distributed between 50\% and 75\% during the training process, while the training logs of other hospitals show that the accuracy of the validation set is almost above 90\% after each round of local training.
However, we have not yet dealt with the weights of participants with poor performance in the training. 
Our next work is to consider constructing an appropriate incentive mechanism to weaken the influence of participants with poor contribution on the overall prediction effect, so as to fully reduce the influence of non-IID data on the model weight.

Furthermore, in order to test the influence of different views on the model prediction effect, as shown in Fig~\ref{fig:03}, we visualize the data of each view. 
We find that the distribution of Spec. is too scattered, and it is difficult to distinguish normal people from patients in special operations such as backspace, space, and keyboard switching. 
Alph. and Accel. have better categorizable results from a single view. This also illustrates from the other hand that there are obvious differences in typing patterns between normal people and depressed patients, including the duration of keystrokes. 
In summary, it is necessary to merge data from different views as input.

\section{Conclusion}\label{sec:conclu}
Federated learning plays a key role in solving the problem of data islands. 
In the problem of data islands, user privacy and data security are very important. 
In this work, we use the data records generated by the user when typing on the mobile phone and the user's HDRS score to predict depression through the DeepMood architecture. 
For IID data, with different amounts of data, the accuracy of federated learning is about 10\%-15\% higher than that of local training without weight update. 
For non-IID data, accuracy is only reduced by 13\% at most. From the perspective of protecting user privacy, this is completely acceptable.

%
%

%
%
\bibliography{ref}

\end{document}